\documentclass[preprint,showpacs,aps,pra,floatfix]{revtex4}
\usepackage{amssymb}

\usepackage{graphicx}
\usepackage{dcolumn}
\usepackage{bm}
\usepackage{setspace}
\usepackage{amsmath}

\providecommand{\U}[1]{\protect\rule{.1in}{.1in}}
\providecommand{\U}[1]{\protect\rule{.1in}{.1in}}
\input{tcilatex}

\begin{document}

\title{Cross section for double charmonium production in electron-positron
annihilation at energy $\sqrt{s}$= 10.6 GeV}
\author{Elias Mengesha, Shashank Bhatnagar}

\begin{abstract}
In this work we study the process $e^{+}+e^{-}\longrightarrow
J/\Psi+\eta_{c} $ at energy $\sqrt{s}= 10.6 GeV$ observed recently at
B-factories whose measurements were made by Babar and Belle groups. We
calculate the cross section for this process in the Bethe-salpeter formalism
under Covariant Instantaneous Anstaz (CIA). To simplify our calculation, the
heavy quark approximation is employed in the quark and gluon propagators. In
the exclusive process of $e^{+}e^{-}$ annihilation into two heavy quarkonia,
the cross section calculated in this scenario is compatible with the
experimental data of Babar and Belle.
\end{abstract}

\maketitle

\affiliation{Department of Physics, Addis Ababa University, P.O.Box 1176, Addis Ababa,
Ethiopia}


12.39.-x, 11.10.St , 21.30.Fe , 12.40.Yx , 13.20.-v 

\section{Introduction}

\label{sec:1}

In this work we study the exclusive production process $e^{-}+
e^{+}\rightarrow J/\psi + \eta_{c}$ at energy $\sqrt{s}=10.6GeV$ observed at
B-factories [1,2,3] whose measurements have recently been done by Babar and
Belle groups. It is well known that there was a significant discrepancy
between the experimental measurements \cite{1,2,3} and the non-relativistic
QCD (NRQCD)\cite{4,5} predictions for this process at centre of mass
energies $\sqrt{s}\approx10.6GeV$. This process has been recently studied in
a Bethe-Salpeter formalism \cite{6} in Instantaneous Approximation (IA). To
simplify calculations, the authors have employed heavy quark limit in the
propagators for studying systems composed of heavy charm and anti-charm
quarks.

We wish to mention that Bethe-Salpeter equation (BSE) is a conventional
non-perturbative approach in dealing with relativistic bound state problems
in QCD. It is firmly established in the framework of field theory and from
the solutions we can obtain useful information about the inner structure of
hadrons, which is also crucial in treating high energy hadronic scattering
and production processes. Despite its drawback of having to input
model-dependent kernel, these studies have become an interesting topic in
recent years, since calculations have shown that BSE framework using
phenomenological potentials can give satisfactory results as more and more
data is being accumulated. Further by adopting this framework we get more
insight about the treatment of this process. This is mainly due to the
unambiguous definition of BS wave function which is expressible by time
ordered product of Heisenberg picture operators. This provides exact
effective coupling vertex for bound state particle with all its N (N = 2 for
mesons) constituents and can be considered as summing up all the
non-perturbative QCD effects in the bound state.

On lines of \cite{6}, we try to study this process in the framework of BSE
under Covariant Instantaneous Ansatz (CIA) which is a Lorentz- invariant
generalization of Instantaneous Approximation (IA). What distinguishes CIA
from the other 3D reductions of BSE is its capacity for a two-way
connection: an exact 3D BSE reduction for a $\mathrm{q}\overline{\mathrm{q}}$
system, and an equally exact reconstruction of original 4D BSE, the former
to make contact with the mass spectrum, while the latter for calculation of
transition amplitudes as 4D quark-loop integrals \cite{mitra01, a, a1,
bhatnagar06, bhatnagar09}. We wish to emphasise here that in these studies
one of the main ingredients is the Dirac structure of the Bethe-Salpeter
wave function (BSW). The copious Dirac structure of BSW was already studied
by Llewllyn Smith \cite{smith69} much earlier. Recent studies \cite
{cvetic04,alkofer02} have revealed that various covariant structures in BSWs
of various hadrons are necessary to obtain quantitatively accurate
observables. It has been further noticed that all covariants do not
contribute equally for calculation of meson observables. To address this
problem, recently we thought of investigating how to arrange these
covariants systematically in BSWs. Thus, in a recent work \cite{bhatnagar06}%
, we developed a power counting rule for incorporating various Dirac
structures in BSW, order-by-order in powers of inverse of meson mass.
According to this power counting rule,the Leading order (LO) covariants are
expected to contribute maximum to calculation of any meson observable,
followed by the next-to-leading order (NLO) covariants. Taking in view of
this fact, as a first step we have outlined the Dirac covariants and
expanded the coefficients to the leading order (LO), and calculated the
leptonic decay constants of vector mesons \cite{bhatnagar06} as well as
pseudoscalar mesons \cite{bhatnagar09} at this order. The results were found
to be close to data. In another recent work \cite{bhatnagar11}, we studied
leptonic decay constants of unequal mass pseudoscalar mesons like $\pi, K,
D, D_{s}$and $B$ and radiative decays of equal mass pseudoscalar mesons like 
$\pi^{0}, \eta_{c}$ by taking into account both the leading order (LO) and
Next-to-Leading Order (NLO) Dirac covariants. It was found that the
contribution of leading order (LO) covariants to decay constants was maximum
(about 90-95 percent) for heavier mesons composed of c and b quarks like $D$,%
$D_{s}$,$B$ and $\eta_{c}$ \cite{bhatnagar11} , while there was little
contribution from NLO covariants. We now calculate the cross section for the
process $e^{-}+ e^{+}\rightarrow J/\psi + \eta_{c}$ by employing the most
leading of the LO covariants (such as $\gamma_{5}$ for heavy pseudoscalar
mesons like $\eta_{c}$ and $i\gamma.\varepsilon$ for heavy vector meson like 
$J/\psi$ comprising of heavy charm and anti-charm quarks for which BS
formalism is quite suitable. In order to simplify the calculation, we will
further impose the heavy quark approximation ($P\sim M, q<<M$) on the quark
and gluon propagators to simplify the integrals as in Ref.\cite{6}. Under
this approximation, our results are comparable with the data \cite{1,2,3}.
The remainder of this paper is organized as follows: In sec.II, we will
study the BS equations for vector and pseudoscalar quarkonia. In sec.III, we
will calculate the amplitude and cross section for the process $%
e^{+}+e^{-}\longrightarrow J/\Psi+ \eta_{c}$ in the BS formalism. Sec.IV is
reserved for conclusions and discussions.

\section{The Bethe-Salpeter Wave Function under CIA}

\label{sec:2}

We briefly outline the BSE framework under CIA. For simplicity, lets
consider a $\mathrm{q}\overline{\mathrm{q}}$ system comprising of scalar
quarks with an effective kernel $K$, 4D wave function $\Phi(P,q)$, and with
the 4D BSE, 
\begin{equation}
i(2\pi)^{4}\Delta_{1}\Delta_{2}\Phi(P,q)=\int
d^{4}q^{\prime}K(q,q^{\prime})\Phi(P,q^{\prime}),  \label{eq:2.1}
\end{equation}
where $\Delta_{1,2}=m_{1,2}^{2}+p_{1,2}^{2}$are the inverse propagators, and 
$m_{1,2}$ are (effective) constituent masses of quarks. The 4-momenta of the
quark and anti-quark, $p_{1,2}$, are related to the internal 4-momentum $%
q_{\mu}$ and total momentum $P_{\mu}$ of hadron of mass $M$ as $%
p_{1,2}{}_{\mu}=\widehat{m}_{1,2}P_{\mu}\pm q_{\mu},$ where $\widehat{m}%
_{1,2}=[1\pm(m_{1}^{2}-m_{2}^{2})/M^{2}]/2$ are the Wightman-Garding (WG)
definitions of masses of individual quarks. Now it is convenient to express
the internal momentum of the hadron $q$ as the sum of two parts, the
transverse component, $\hat{q}_{\mu}=q_{\mu}-\frac{q.P}{P^{2}}P_{\mu}$ which
is orthogonal to total hadron momentum $P$ (ie. $\widehat{q}.P=0$ regardless
of whether the individual quarks are on-shell or off-shell), and the
longitudinal component, $\sigma P_{\mu}=(q\cdot P/P^{2})P_{\mu}$, which is
parallel to P. To obtain Hadron-quark vertex, use an Ansatz on the BS kernel 
$K$ in Eq. (\ref{eq:2.1}) which is assumed to depend on the 3D variables $%
\hat{q}_{\mu}$, $\hat{q}_{\mu}^{\prime}$ \cite{mitra01, a, bhatnagar06,
bhatnagar09, bhatnagar11} i.e.

\begin{equation}
K(q,q^{\prime})=K(\hat{q},\hat {q}^{\prime}),
\end{equation}

(A similar form of the BS kernel was also earlier suggested in ref. \cite
{resag94}). Hence, the longitudinal component, $\sigma P_{\mu }$ of $q_{\mu }
$, does not appear in the form $K(\hat{q},\hat{q}^{\prime })$ of the kernel.
For reducing Eq.(1) to the 3D form of BSE, we define a 3D wave function $%
\phi (\widehat{q})$as,

\begin{equation}
\phi(\hat{q})=\int^{+\infty}_{-\infty}Md\sigma\Phi(P,q)
\end{equation}

Substituting Eq.(3) in Eq.(1), with the definition of the kernel in Eq.(2),
we get a covariant version of the Salpeter equation which is in fact a 3D
BSE:

\begin{equation}
(2\pi)^{3}D(\hat{q})\phi(\hat{q})=\int d^{3}\hat{q}^{\prime}K(\hat{q},\hat{q}%
^{\prime})\phi(\hat{q}^{\prime}).
\end{equation}

Here $D(\hat{q})$ is the 3D denominator function defined below whose value
is obtained by evaluating contour integration over inverse quark propagators
in the complex $\sigma$-plane by noting their corresponding pole positions 
\cite{bhatnagar05,bhatnagar06,bhatnagar09}as, 
\begin{equation}
\frac{1}{D(\hat{q})}=\frac{1}{2\pi i}\int\limits_{-\infty}^{+\infty}\frac{%
Md\sigma}{\Delta_{1}\Delta_{2}}=\frac{\frac{1}{2\omega_{1}}+\frac
{1}{2\omega_{2}}}{(\omega_{1}+\omega_{2})^{2}-M^{2}}%
;\omega_{1,2}^{2}=m_{1,2}^{2}+\hat{q}^{2}.  \label{eq:2.5}
\end{equation}
We note that the RHS of Eq.(4) is exactly identical to the RHS of Eq.(1) by
virtue of Eq.(2) and Eq.(3). We thus have an exact interconnection between
3D BSE and the 4D BSE, and hence between the 3D wave function$\phi(\hat{q})$
and the 4D wave function$\Phi(P,q)$\cite
{mitra01,a,a1,bhatnagar06,bhatnagar09}: 
\begin{equation}
\Delta_{1}\Delta_{2}\Phi(P,q)=\frac{D(\hat{q})\phi(\hat{q})}{2\pi i}%
\equiv\Gamma(\hat{q}),  \label{eq:2.7}
\end{equation}
where $\Gamma(\hat{q})$ is the Bethe-Salpeter Hadron-quark vertex function
for a meson comprising of scalar quarks.

The 4D BS wave function $\Phi(P,q)$ can be reconstructed from the 3D BS wave
function $\phi(\hat{q})$ as: 
\begin{equation}
\Phi(P,q)=\frac{1}{\Delta_{1}}\Gamma(\hat{q})\frac{1}{\Delta_{2}},
\end{equation}
where $\Delta_{i}=(m_{i}^{2}+p_{i}^{2})$,(i=1,2) are the inverse propagators
for scalar quarks which flank the hadron-quark vertex$\Gamma$. This 4D
hadron-quark vertex $\Gamma(\hat{q})$ satisfies a 4D BSE with a natural
off-shell extension over the entire 4D space (due to the positive
definiteness of the quantity $\hat{q}^{2}=q^{2}-(q\cdot P)^{2}/P^{2}$
throughout the entire 4D space) and thus provides a fully Lorentz-invariant
basis for evaluation of various transition amplitudes through various quark
loop diagrams. Due to these properties, this framework of BSE under CIA can
be profitably employed not only for low energy studies but also for
evaluation of various transition amplitudes at quark level all the way from
low energies to high energies. However this 4D hadron-quark vertex $\Gamma$
is still unnormalized and can be normalized as will be shown in the
realistic case of fermionic quarks next.

For fermionic quarks the BSE under CIA can be written as:\bigskip 
\begin{equation}
i(2\pi)^{4}\Psi(P,q)=S_{F1}(p_{1})S_{F2}(p_{2})\int d^{4}q^{\prime}K(%
\widehat{q},\widehat{q}^{\prime})\Psi(P,q^{\prime}),
\end{equation}

where the scalar propagators $\Delta_{i}^{-1}$ in the above equations are
replaced by fermionic propagators $S_{F}$. Further the $H{q\bar{q}}$ vertex
would be a $4\times4$ matrix in the spinor space for which we should
incorporate the relevant Dirac structures. For incorporation of the relevant
Dirac structures in $\Gamma(\hat{q})$, they are incorporated order-by-order
in powers of inverse of meson mass $M$, in accordance with the power
counting rule we developed in \cite{bhatnagar06}. Our aim of developing the
power counting rule was to find a ``criterion'' so as to systematically
choose among various Dirac covariants from their complete set to write wave
functions for different mesons (vector mesons, pseudoscalar mesons etc.)\cite
{bhatnagar06,bhatnagar09}. In another recent work, we studied leptonic decay
constants of unequal mass pseudoscalar mesons like $\pi, K, D, D_{s}and B$
and radiative decays of equal mass pseudoscalar mesons like $\pi^{0},
\eta_{c}$ by taking into account both the leading order (LO) and
Next-to-Leading Order (NLO) Dirac covariants. It was found that the
contribution of leading order (LO) covariants to decay constants was maximum
(about 90-95 percent) for heavier mesons composed of c and b quarks like $D$,%
$D_{s}$,$B$and $\eta_{c}$\cite{bhatnagar11}, while there was little
contribution from NLO covariants. Among the LO covariants, it was also
noticed that for pseudoscalar mesons, the contribution from covariant, $%
\gamma_{5}$was maximum and similarly for vector mesons, the contribution of
LO covariant $i\gamma.\varepsilon$was maximum.

Thus to simplify the calculations, as a first step, we calculate the cross
section for the process $e^{-}+ e^{+}\rightarrow J/\psi + \eta_{c}$by
employing the most leading of the LO covariants such as $\gamma_{5}$for
heavy pseudoscalar mesons like $\eta_{c}$and $i\gamma.\varepsilon$for heavy
vector meson like $J/\psi$comprising of heavy charm and anti-charm quarks
for which BS formalism is quite suitable. The full fledged normalized 4D BS
wave functions for a $q\bar{q}$ meson with quarks total momentum $P$and
relative momentum $q$ and with individual quarks with momenta $p_{1}$ and $%
p_{2}$ can be written as:

\begin{equation}
\Psi(P,q)=S_{F}(p_{1})\Gamma(\hat{q})S_{F}(-p_{2})
\end{equation}
where the 4D BS hadron-quark vertex function which absorbs the 4D BS
normalizer N is, 
\begin{equation}
\Gamma(\hat{q})=N\Gamma_{i}D(\hat{q})\phi(\hat{q})/2\pi i
\end{equation}
where $\Gamma_{i}=\gamma_{5},i\gamma.\varepsilon,...$ are the relevant Dirac
structures for pseudoscalar mesons, vector mesons etc. The 4D BS normalizer $%
N$ which is determined from standard current conserving conditions and is
worked out in the framework of Covariant Instantaneous Ansatz (CIA) to give
explicit covariance to the full fledged 4D BS wave function, $\Psi(P,q)$ and
hence to the Hadron-quark vertex function, $\Gamma(\hat{q})$ employed for
calculation of transition amplitudes at high energies.

Thus the 4D hadron-quark vertex function for $\eta_{c}$ is, 
\begin{equation}
\Gamma^{P}(\widehat{q_{b}})=\gamma_{5}N_{P}D(\widehat{q_{b}})\phi(\widehat{%
q_{b}})/2\pi i,
\end{equation}

while the 4D hadron-quark vertex function for $J/\Psi$ is, 
\begin{equation}
\Gamma^{V}(\widehat{q_{a}})=i\gamma.\varepsilon N_{V}D(\widehat{q_{a}})\phi(%
\widehat{q_{a}})/2\pi i.
\end{equation}

Here $N_{P}$ and $N_{V}$ are the 4D BS normalizers for $\eta_{c}$ and $%
J/\Psi $ meson (with internal momenta $q_{a}$ and $q_{b}$ respectively)
which are determined through the current conservation condition \cite
{mitra01,a,bhatnagar06,bhatnagar09,bhatnagar11}, while $D(\widehat{q_{a}})$
and $D(\widehat{q_{b}})$ are the respective denominator functions. $\phi(%
\widehat{q_{a}})$ and $\phi(\widehat{q_{b}})$ are the 3D BS wave functions
for $\eta_{c}$ and $J/\psi$ respectively, while $\varepsilon$ is the
polarization vector for the $J/\psi$ meson.

As far as the input kernel $K(q,q^{\prime})$\cite
{mitra01,a,bhatnagar06,bhatnagar09,bhatnagar11}, in BSE is concerned, it is
taken as one-gluon-exchange like as regards color [$(\bm{\lambda}%
^{(1)}/2)\cdot(\bm{\lambda}^{(2)}/2)$] and spin ($\gamma_{\mu}^{(1)}\gamma_{%
\mu }^{(2)}$) dependence. The scalar function $V(q-q^{\prime})$ is a sum of
one-gluon exchange $V_{OGE}$ and a confining term $V_{conf}$. Thus we can
write the interaction kernel as \cite{bhatnagar06,mitra01}: 
\begin{equation*}
K(q,q^{\prime})=\left( \frac{1}{2}\bm{\lambda}^{(1)}\right) \cdot\left( 
\frac{1}{2}\bm{\lambda}^{(2)}\right) V_{\mu}^{(1)}V_{\mu}^{(2)}V(q-q^{\prime
});
\end{equation*}
\begin{equation*}
V_{\mu}^{(1,2)}=\pm2m_{1,2}\gamma_{\mu}^{(1,2)};
\end{equation*}

\begin{equation}
\begin{array}{rcl}
V(\hat{q}-\hat{q}^{\prime}) & = & \displaystyle\frac{4\pi\alpha_{S}(Q^{2})}{(%
\hat{q}-\hat{q}^{\prime})^{2}}+\frac{3}{4}\omega_{q\bar{q}}^{2}\int d^{3}%
\bm{r}\left[ r^{2}(1+4a_{0}\hat{m}_{1}\hat{m}_{2}M^{2}r^{2})^{-1/2}-\frac{%
C_{0}}{\omega_{0}^{2}}\right] e^{i(\bm{\hat{q}-\hat{q}'})\cdot \bm{r}}; \\ 
&  &  \\ 
\alpha_{S}(Q^{2}) & = & \displaystyle\frac{12\pi}{33-2f}\left( \ln\frac
{Q^{2}}{\Lambda^{2}}\right) ^{-1}.
\end{array}
\label{eq:2.14}
\end{equation}
The Ansatz employed for the spring constant $\omega_{q\overline{q}}^{2}$ in
Eq. (\ref{eq:2.14}) is \cite{mitra01,a,bhatnagar06,bhatnagar09,bhatnagar11}, 
\begin{equation}
\omega_{q\overline{q}}^{2}=4\widehat{m}_{1}\widehat{m}_{2}M_{>}%
\omega_{0}^{2}\alpha_{S}(M_{>}^{2}), \quad M_{>}=Max(M,m_{1}+m_{2})
\label{eq:2.15}
\end{equation}

Here $\widehat{m}_{1}$, $\widehat{m}_{2}$ are the Wightman-Garding
definitions of masses of constituent quarks defined earlier. Here the
proportionality of $\omega_{q\bar{q}}^{2}$ on $\alpha_{S}(Q^{2})$ is needed
to provide a more direct QCD motivation to confinement. This assumption
further facilitates a flavour variation in $\omega_{q\bar{q}}^{2}$. And $%
\omega _{0}^{2}$ in Eq. (\ref{eq:2.14}) and Eq. (\ref{eq:2.15}) is
postulated as a universal spring constant which is common to all flavours.
Here in the expression for $V(\hat{q}-\hat{q}^{\prime})$, as far as the
integrand of the confining term $V_{conf}$ is concerned, the constant term $%
C_{0}/\omega _{0}^{2}$ is designed to take account of the correct zero point
energies, while $a_{0}$ term ($a_{0}\ll1$) simulates an effect of an almost
linear confinement for heavy quark sectors (large $m_{1}$, $m_{2}$), while
retaining the harmonic form for light quark sectors (small $m_{1}$, $m_{2}$) 
\cite{mitra01} as is believed to be true for QCD. Hence the term $%
r^{2}(1+4a_{0}\widehat{m}_{1}\widehat{m}_{2}M_{>}^{2}r^{2})^{-1/2}$ in the
above expression is responsible for effecting a smooth transition from
harmonic ($q\overline{q}$) to linear ($Q\overline{Q}$) confinement . The
values of basic constants are: $C_{0}=.29,\ \omega_{0}=0.158$ GeV, $%
m_{u,d}=0.265$ GeV, $m_{s}=0.415$ GeV, $m_{c}=1.530$ GeV and $m_{b}=4.900$
GeV which have been earlier fit to the mass spectrum of $q\overline{q}$
mesons\cite{mitra01} obtained by solving the 3D BSE under Null-Plane Ansatz
(NPA). However due to the fact that the 3D BSE under CIA has a structure
which is formally equivalent to the 3D BSE under NPA, near the surface $%
P.q=0 $, the $q\overline{q}$ mass spectral results in CIA formalism are
exactly the same as the corresponding results under NPA formalism\cite
{mitra01,a,bhatnagar05}. The details of BS model under CIA in respect of
spectroscopy are thus directly taken over from NPA formalism (For details,
see \cite{mitra01,a, bhatnagar05}). As far as the 3D wave function $\phi(%
\hat{q})$ is concerned, it satisfies the 3D BSE on the surface P.q=0, which
is appropriate for making contact with the mass spectra \cite{mitra01}. Its
fuller structure is reducible to that of a 3D harmonic oscillator. The
ground state wave function deducible from this equation has a gaussian
structure and is expressible as $\phi(\hat{q})\approx e^{-\hat{q}%
^{2}/2\beta^{2}}$, where $\beta$ is the inverse range parameter which
incorporates the content of BS dynamics and is dependent on the input kernel
and is given as in \cite{mitra01,bhatnagar06,bhatnagar09}). The structure of
inverse range parameter $\beta$ in wave function $\phi(\widehat{q})$ is
given as\cite{mitra01,bhatnagar06,bhatnagar09}:

\begin{equation}
\beta^{2}=(2\widehat{m}_{1}\widehat{m}_{2}M\omega_{q\bar{q}%
}^{2}/\gamma^{2})^{1/2}; \gamma^{2}=1-\frac{2\omega_{q\bar{q}}^{2}C_{0}}{%
M_{>}\omega_{0}^{2}}
\end{equation}

We now give the calculation of amplitude and cross section for the process $%
e^{-}+ e^{+}\rightarrow J/\psi + \eta_{c}$ the leading order (LO) of QCD in
the next section. We employ the hadron-quark vertex functions for $\eta_{c}$
and $J/\psi$ mesons given in Eq.(11) and (12) respectively in this
calculation.

\section{Calculation of amplitude and cross section for the process $%
e^{+}+e^{-}\longrightarrow J/\Psi+\protect\eta_{c}$}

\label{sec:3}

There are four Feynman diagrams in the leading order (LO) of QCD for the
process $e^{+}+e^{-}\longrightarrow J/\Psi +\eta _{c}$. Two of these are
depicted in Fig.1. The other two diagrams can be obtained by permutations.
The details of momentum labeling of the diagram in Fig.1a is shown in Fig. 2
below. With reference to the momentum labeling in Fig.2, the adjoint BS wave
function for $\eta _{c}$ meson can be written down as: 
\begin{equation}
\overline{\Psi }(P_{b},q_{b})=S_{F}(-q_{2})\Gamma ^{P}(\widehat{q_{b}}%
)S_{F}(q_{4})
\end{equation}
, while for $J/\psi $ meson, the adjoint BS wave function can be written as: 
\begin{equation}
\overline{\Psi }(P_{a},q_{a})=S_{F}(-q_{3})\Gamma ^{V}(\widehat{q_{a}}%
)S_{F}(q_{1}^{\prime })
\end{equation}
where $q_{a},q_{b}$ are the internal momenta of the hadrons $J/\Psi $ and $%
\eta _{c}$ respectively with the corresponding hadron-quark vertex functions 
$\Gamma ^{V}$ and $\Gamma ^{P}$ given in Eq.(11-12). 
\begin{figure}[h]
\centering
\includegraphics[width=12cm,height=4cm]{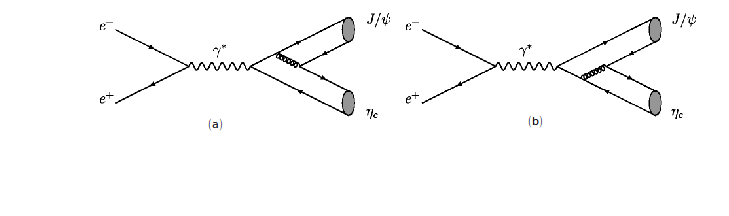}\newline
\caption{Two of the lowest order Feynman diagrams for the production of a
pair of doubly heavy $c\overline{c}$ mesons in $e^{+}e^{-}$ annihilation.
Other two diagrams can be obtained by permutations.}
\label{2}
\end{figure}

\begin{figure}[h]
\centering
\includegraphics[width=14cm,height=5cm]{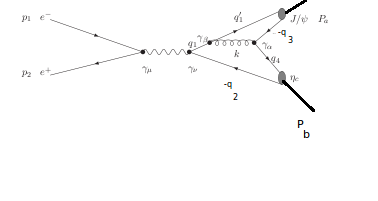}\newline
\caption{Momentum labeling of the first of the two Feynman diagrams shown in
Fig.1}
\label{2}
\end{figure}
Using Feynman rules, one can obtain the amplitude for each of the diagrams
in Fig.1. The amplitude corresponding to process in Fig.1a (described in
detail in Fig.2) is given by 
\begin{equation}
M_{1}=c\delta_{\mu\nu}ee_{Q}g_{s}^{2}\frac{1}{s}\overline{v}%
(p_{2})\gamma_{\mu}u(p_{1})\int d^{4}q_{a}d^{4}q_{b}Tr[\overline{\Psi}%
(P_{a},q_{a})\gamma_{\beta}S_{F}(q_{1})\gamma_{\nu}\overline{\Psi}%
(P_{b},q_{b})\gamma_{\alpha}]\frac{\delta_{\alpha\beta}}{k^{2}}
\end{equation}
which can in turn be expressed as, 
\begin{eqnarray}
M_{1}=&& \frac{c\delta_{\mu \nu} e e_{Q}g_{s}^{2}}{s}\overline{v}%
(p_{2})\gamma_{\mu}u(p_{1})\int d^{4}q_{a}d^{4}q_{b}  \notag \\
&& \times Tr[S_{F}(-q_{3})\Gamma_{v}(\hat{q_{a}})S_{F}(q^{\prime}_{1})%
\gamma_{\beta}S_{F}(q_{1})\gamma_{\nu}S_{F}(-q_{2})\Gamma_{p}(\hat{q_{b}}%
)S_{F}(q_{4})\gamma_{\alpha}]\frac{\delta_{\alpha\beta}}{k^{2}}  \notag \\
\end{eqnarray}

where c =$\frac{4}{3}$ is the color factor, the Mendelstam variable s is
defined as, $s=-(p_{1}+p_{2})^{2}$ and $e_{Q}=2e/3$ is the electric charge
of the charmed quark. The momentum relations of the quark and anti-quark in
the final state are: 
\begin{equation}
q_{1}^{\prime }=\frac{1}{2}P_{a}+q_{a},q_{3}=\frac{1}{2}P_{a}-q_{a},q_{4}=%
\frac{1}{2}P_{b}+q_{b},q_{2}=\frac{1}{2}P_{b}-q_{b},
\end{equation}
and the momenta in the gluon and the quark propagators are given by 
\begin{equation}
k=q_{3}+q_{4}=\frac{1}{2}(P_{a}+P_{b})-q_{a}+q_{b},
\end{equation}
\begin{equation}
q_{1}=q_{1}^{\prime }+k=P_{a}+\frac{1}{2}P_{b}+q_{b},
\end{equation}
respectively. As each of quark momenta in the quark propagators as well as
the gluon propagator is going to depend upon the internal hadron momenta $%
q_{a}$ and $q_{b}$, the calculation of amplitude is going to involve
integrations over these internal momenta and will be quite complex. Hence
following \cite{6}, we simplify the calculation, by employing the heavy
quark approximation on the quark propagators, where we take the quark masses
to be much larger than the internal momenta $q_{a}$ and $q_{b}$ of the
hadrons. In this heavy quark approximation,we can use the approximation, $%
q_{a}<<M_{a},q_{b}<<M_{b}$. Thus we can write, 
\begin{equation}
q_{a(b)}\ll P_{a(b)}\sim M_{a(b)},k=\frac{1}{2}(P_{a}+P_{b})-q_{a}+q_{b}%
\approx \frac{1}{2}(P_{a}+P_{b}),q_{1}=P_{a}+\frac{1}{2}P_{b}+q_{b}\approx
P_{a}+\frac{1}{2}P_{b}.
\end{equation}
With the above approximation, $k^{2}$ and $q_{1}^{2}$ are given by $%
k^{2}\approx -\frac{s}{4}$ and $q_{1}^{2}\approx -\frac{s}{2}-m_{c}^{2}.$
The propagators for the quarks and anti-quarks in momentum space in
Eq.(16-17) are given by $S_{F}(q_{i})=\frac{-i}{i\gamma .q_{i}+m_{c}}$ =$%
\frac{-i(-i\gamma .q_{i}+m_{c})}{\Delta _{i}}$, where index $i$ labels the
quark in the diagram.

Using Eq.(16) and (17) and the preceding expressions for the gluon and the
quark propagators, the amplitude $M_{1}$ in Eq.(19) can be written as:

\begin{equation}
M_{1}= \frac{-2^{6}}{3^{2}s^{3}}g_{s}^{2}e^{2}\overline{v}%
(p_{2})\gamma_{\mu}u(p_{1})\int d^{4}q_{a} d^{4}q_{b}\frac{1}{%
\Delta_{2}\Delta_{3}\Delta_{4}\Delta_{5}} [TR]\frac{N_{v}D_{v}(\hat{q_{a}}%
)\phi_{v}(\hat{q_{a}})}{2\pi i}\frac{N_{p}D_{p}(\hat{q_{b})}\phi_{p}(\hat{%
q_{b}})}{2\pi i} \\
.
\end{equation}

Here [TR] is the trace over the gamma matrices appearing in the quark
propagators in Eq.(19). Noting that the 4-dimensional volume element $%
d^{4}q=d^{3}\hat{q}Md\sigma$, we then perform contour integrations in the
complex $\sigma$-plane by making use of the corresponding pole positions 
\cite{bhatnagar05,bhatnagar09}. The pole integrations over $%
dq_{a}^{0}=M_{a}d\sigma_{a}$ and $dq_{b}^{0}=M_{b}d\sigma_{b}$ in Eq.(24)
can be expressed as:$\frac{1}{2\pi i}\int_{-\infty}^{\infty}\frac{%
M_{a}d\sigma_{a}}{\Delta_{3}\Delta_{5}}=\frac{1}{D(\hat{q_{a}})}$ and $\frac{%
1}{2\pi i}\int_{-\infty}^{\infty}\frac{M_{b}d\sigma_{b}}{\Delta_{2}\Delta_{4}%
}=\frac{1}{D(\hat{q_{b}})}$, where values of denominator functions $D(\hat{q}%
_{a})$ and $D(\hat{q}_{b})$ evaluated by contour integration in the complex $%
\sigma$- plane are expressible as in Eq.(5). After calculating the trace
part in the above equation and employing the heavy quark approximation on
relative momenta given in Eq.(23), one obtains: 
\begin{equation}
M_{1}=\frac{-2^{14}\pi^{2}\alpha_{em}\alpha_{s}m_{c}^{3}}{3^{2}s^{3}}%
\epsilon_{\alpha\mu\lambda\sigma}\varepsilon_{\alpha}P_{b\lambda}P_{a\sigma}%
\int d^{3}\hat{q_{a}}d^{3}\hat{q_{b}}N_{v}N_{p}\phi_{v}(\hat{q_{a}})\phi_{p}(%
\hat{q_{b}})[\overline{v}(p_{2})\gamma_{\mu}u(p_{1})]
\end{equation}
where $\alpha_{em}=\frac{e^{2}}{4\pi}$, while $\alpha_{s}=g_{s}^{2}/4\pi$
and is given in Eq.(13). Let's define $\xi_{a}=\int d^{3}\hat{q_{a}}%
N_{v}\phi_{v}(\hat{q_{a}})$ and $\xi_{b}=\int d^{3}\hat{q_{b}}N_{p}\phi_{p}(%
\hat{q_{b}})$, which are values of wave functions at origins of $J/\psi$ and 
$\eta_{c}$ respectively.\newline
Thus we can express the amplitude $M_{1}$ as, 
\begin{equation}
M_{1}=\frac{-2^{14}\pi^{2}\alpha_{em}\alpha_{s}m_{c}^{3}}{3^{2}s^{3}}%
\epsilon_{\alpha\mu\lambda\sigma}\varepsilon_{\alpha}P_{b\lambda}P_{a\sigma}[%
\overline{v}(p_{2})\gamma_{\mu}u(p_{1})]\xi_{a}\xi_{b}
\end{equation}
Here $N_{p}$ and $N_{v}$ in $\xi_{a,b}$ are the BS normalizers for $\eta_{c}$
and $J/\psi$ respectively which are evaluated by using the current
conservation condition, 
\begin{equation}
2iP_{\mu}=(2\pi)^{4}\int d^{4}q Tr[\overline{\psi}(P,q)(\frac{\partial}{%
\partial P_{\mu}}S_{F}^{-1}(p_{1}))\psi(P,q)S_{F}^{-1}(-p_{2})] +
(1\leftrightarrows 2)
\end{equation}

Putting BS wave function $\psi(P,q)$ for a given meson in the above
equation,carrying out derivatives of inverse quark propagators of
constituent quarks with respect to total hadron momentum $P_{\mu}$,
evaluating trace over gamma matrices, following usual steps and multiplying
both sides of the equation by $P_{\mu}/(-M^{2})$ to extract out the
normalizer N from the above equation, we then express the above expression
in terms of integration variables $\hat{q}$ and $\sigma$. Noting that the
4-dimensional volume element $d^{4}q=d^{3}\hat{q}Md\sigma$, we then perform
contour integration in the complex $\sigma$-plane by making use of the
corresponding pole positions. For details of these mathematical steps
involved in the calculation of BS normalizers for vector and pseudoscalar
mesons, see Ref.\cite{bhatnagar06}and Ref.\cite{bhatnagar11} respectively,
where in the present calculation we take only the leading order Dirac
covariants $i\gamma.\varepsilon$ and $\gamma_{5}$ for $J/\psi$ and $\eta_{c}$
respectively in their respective 4D BS wave functions $\Psi(P,q)$. Then
numerical integration on variable $\hat{q} $ is performed. The values of BS
normalizers thus obtained for $J/\psi$ and $\eta$ mesons are $N_{v}=.0504
GeV^{-3}$ and $N_{P}=.0410GeV^{-3}$ respectively.

The total amplitude for the process $e^{+}e^{-}\longrightarrow
J/\Psi\eta_{c} $ can be obtained by summing over the amplitudes of all the
four diagrams shown in Fig.1. For that matter the amplitude obtained from
the first diagram is the same as the amplitude from each of the remaining
three diagrams in Fig.1. Thus, the total amplitude is 4 times the amplitude
from the first diagram. The unpolarized total cross section is obtained by
summing over various $J/\Psi$ spin-states and averaging over those of the
initial state $e^{+}e^{-}$. Thus, in the CM frame the total cross section, $%
\sigma$, is given by 
\begin{equation}
\sigma=\frac{4m_{e}^{2}}{32\pi}\frac{\mid p_{f}\mid}{\mid
p_{i}\mid(E_{1}+E_{2})^{2}}\int \frac{1}{4}\sum_{spin}\mid
M_{tot}\mid^{2}dcos\theta
\end{equation}
where $p_{f}$ is the momentum of either of the outgoing particles and $p_{i}$
is the momentum of either of the ingoing particles, which is in turn
expressible as, 
\begin{equation}
\sigma=\frac{4m_{e}^{2}}{32\pi}\frac{\sqrt{s-16m_{c}^{2}}}{s^{\frac{3}{2}}}%
\int \frac{1}{4}\sum_{spin}\mid M_{tot}\mid^{2}dcos\theta
\end{equation}
where the masses of the leptons are ignored in the calculation. Explicitly $%
\mid M_{tot}\mid^{2}$ is given by 
\begin{equation}
\frac{1}{4}\sum_{spin}\mid M_{tot}\mid^{2}=\frac{2^{30}\pi^{4}%
\alpha_{em}^{2}\alpha_{s}^{2}m_{c}^{6}}{3^{4}s^{5}4m_{e}^{2}}%
(-32m_{c}^{4}+t^{2}+u^{2})\xi_{a}^{2}\xi_{b}^{2}
\end{equation}
where $t=-(p_{1}-P_{a})^{2}$ and $u=-(p_{1}-P_{b})^{2}$ are the Mandelstam's
variables. Therefore, in the CM frame, the total cross section is given by: 
\begin{equation}
\sigma=\frac{2^{30}\pi^{3}\alpha_{em}^{2}\alpha_{s}^{2}m_{c}^{6}}{83^{5}s^{4}%
}(1-\frac{16m_{c}^{2}}{s})^{\frac{3}{2}}\xi_{a}^{2}\xi_{b}^{2}
\end{equation}

\bigskip \textbf{Numerical Results:}\newline
The basic input parameters in the calculation are just four: $C_{0}=0.29$, $%
\omega_{0}=0.158GeV$, QCD length scale, $\Lambda_{QCD}=0.200GeV $, and the
charmed quark mass, $m_{c}=1.530GeV$ \cite{bhatnagar06,bhatnagar09}. The
numerical values of inverse range parameter $\beta$ calculated from Eq.(15)
are $\beta_{J/\psi}=.4989GeV$ and $\beta_{\eta_{c}}=.4388GeV$. To calculate
the values of $\beta$, for the two hadrons, the experimental hadron masses
are taken as, $M_{J/\psi}=3.096GeV$ and $M_{\eta_{c}}=2.982GeV$. With these
parameters the total cross section for the above process at $\sqrt{s}=
10.6GeV$ is calculated to be $\sigma=21.75fb$.\newline

\section{Discussion}

\label{sec:4} In this paper we have calculated the cross section of the
exclusive process of $e^{+}e^{-}\longrightarrow J/\Psi\eta_{c}$ at energy $%
\sqrt{s}=10.6GeV$ in the framework of BSE under CIA \cite
{bhatnagar09,bhatnagar11,bl07} using only the leading order (LO) diagrams in
QCD. We find the theoretical value of $\sigma[e^{+}e^{-}\longrightarrow
J/\Psi\eta_{c}]=21.75fb$, which is broadly in agreement with the Babar's
data $\sigma[e^{+}e^{-}\longrightarrow J/\Psi\eta_{c}]=(17.6\pm2.8\pm2.1)fb$ 
\cite{1} and the Belle's data, $\sigma[e^{+}e^{-}\longrightarrow
J/\Psi\eta_{c}]=(25.6\pm2.8\pm3.4)fb$ \cite{2,3}.

It had been noticed earlier that NRQCD predictions \cite{4,5} for the above
process at $\sqrt{s}= 10.6 GeV$ using leading order diagrams alone give
cross sections which are much less than data \cite{1,2,3}. Such a large
discrepancy between experimental results and theoretical predictions has
been a challenge to the understanding of charmonium production through
NRQCD. Many studies were performed to resolve this problem. For instance
Braaten and Lee\cite{braaten03} first showed that results on cross section
are found to improve considerably when relativistic corrections are
incorporated. Then it was found that to obtain cross sections from NRQCD
which are consistent with data, one has to incorporate NLO QCD corrections 
\cite{gong07,zhang06}. However in these studies it was found that the value
of total NLO contribution to cross section is nearly twice the LO
contribution. In the present calculations under the relativistic framework
of BSE under CIA, we obtained results for cross sections which are in good
agreement with data\cite{1,2,3} using leading order QCD processes alone
though we have employed the heavy quark approximation ($q<<M$ and $P\sim M$)
on the quark and gluon propagators on lines of \cite{6}. However we have not
made use of the non-covariant heavy quark limit as in Eq.(3-4) of Ref.[6]and
work with the exact propagators of the quarks constituting the two hadrons.
This is a validation of the fact that BSE which is firmly rooted in field
theory and which incorporates relativistic effects within its premises is
ideally suited to describe not only low energy processes, but even processes
at high energies such as high energy hadronic scatterings and production
processes.

The approach in this paper is quite different from the approach in Ref.[6]
is the sense that we employ the framework of BSE under CIA which is a
relativistic generalization of Instantaneous Approximation (IA) used in the
former and has a much wider range of applicability as explained in Section
II of this paper. Further, to calculate their results, Ref.[6] has made use
of heavy quark limit on the propagators of all the heavy quarks and
anti-quarks, where the propagators have been simplified as in their Eq.(3)
and (4) which in fact is non-relativistic and non-covariant. Also to
simplify their calculation of amplitude in their Eq.(22), [6] makes use of
the heavy quark approximation, in that the propagators of quark and gluon
are independent of relative momenta $q_{a}$ and $q_{b}$ of the two hadrons
since the masses of quarks are large compared to their relative momentum.
However in our paper, we only make use of the above heavy quark
approximation ($q<<M$ and $P\sim M$) only in the sense of simplifying the
integrals involved in Eq.(18)-(19) for amplitude calculation as done in [6],
but we do not employ the non-relativistic and non-covariant heavy quark
limit on the quark and anti quark propagators (as in Eq.(3) and (4) of [6])
and instead work with the full quark and anti-quark propagators for the
quarks constituting the hadrons. In doing so using CIA, we also see that our
results on $\sum_{s}|M_{tot}|^{2}$ in Eq.(30) and cross section in Eq.(31)
of our paper are not exactly similar to results of [6]. In this regard, we
wish to point out that our amplitude and cross sectional formulae involve
the 4D BS normalizers $N_{V}$ and $N_{P}$ which are calculated in the
framework of CIA and whose numerical values are explicitly worked out for
both the hadrons, $J/\psi $ and $\eta _{c}$ as $N_{v}=.0504GeV^{-3}$ and $%
N_{P}=.0410GeV^{-3}$ respectively. These normalizers enter the amplitude and
cross sectional formulae through the definitions of values of wave functions$%
\xi _{a}$ and $\xi _{b}$ of the two hadrons at their origins, which is not
so in case of Ref.[6]. Further, the input BSE kernel and hence the input
parameters employed by us and by Ref.[6] are completely different. While we
employ ''vector'' confinement, ie. we make use of a common form $(\gamma
_{\mu }^{(1)}\gamma _{\mu }^{(2)})$ for both one-gluon exchange as well as
confining terms in the kernel as in Eq.(13) (see\cite{mitra01,bhatnagar06}%
for details) , Ref.[6] employs a scalar $(1^{(1)}.1^{(1)})$ form for
confinement, while vector form $(\gamma _{\mu }^{(1)}\gamma _{\mu }^{(2)})$
for one-gluon exchange. Further the functional form of confining potential
in [6]is also different from our case. Whereas Ref.[6] uses linear
confinement $(\thicksim r)$ which is true in case of heavy quark (c,b)
systems, we have used a general form of confinement potential in Eq.(13)
which simulates the effect of linear confinement $(\thicksim r)$ for heavy
quark sector (large $m_{1},m_{2})$ while retaining harmonic form $(\thicksim
r^{2})$ for light quark sector (small $m_{1},m_{2})$ as explained in section
2. As far as the numerical results on cross section in this paper and
Ref.[6] are concerned, they are quits close. This may be due to the fact
that for systems comprising of heavy quarks (c,b), the CIA results may lead
to IA results when heavy quark approximation is imposed. However this may
not be so for systems comprising of light quarks (u,d,s).

However in the present calculation, we used only the first of the leading
order(LO) Dirac covariants ($\gamma_{5}$ and $i\gamma.\varepsilon$)in
hadron-vertex functions for $J/\psi$ and $\eta_{c}$ mesons respectively.
They were identified as most leading covariants in accordance with our power
counting scheme. These covariants give maximum contribution to calculation
of meson observables such as decay constants etc. It can also be seen here
that the results on cross section for the process $e^{+}e^{-}\longrightarrow
J/\Psi+\eta_{c}$ employing these most leading of the LO covariants brings
theoretical results close to data \cite{1,2,3}. We now also intend to see
the effect of incorporation of both LO and the NLO Dirac covariants (to the
vertex functions of these mesons) on cross section for the process studied.
The contribution from NLO covariants is expected to be much lesser than the
contribution from LO covariants in line with our recent studies on meson
decays\cite{bhatnagar06,bhatnagar09,bhatnagar11}. And this is more so for
heavy mesons comprising of c and b quarks. It is expected that the results
of cross section will improve further with incorporation of both LO and NLO
covariants and without employing the heavy quark approximation on the quark
and gluon propagators. This calculation will be quite rigorous and will be
the subject of a later communication.

\bigskip

\textbf{Acknowledgements: } This work was carried out in the Department of
Physics, Addis Ababa University (AAU). The authors would like to thank the
Physics Department,AAU for the facilities provided during the course of this
work. One of us, EM would like to thank Haramaya University (HU) for
supporting his doctoral programme.\bigskip

REFERENCES:

\end{document}